\documentstyle[psfig,times]{mn}

\def\lesssim{\hbox{$\;\raise.4ex\hbox{$<$}
                \kern-.75em\lower.7ex\hbox{$\sim$}\;$}}
\def\gtrsim{\hbox{$\;\raise.4ex\hbox{$>$}
                \kern-.75em\lower.7ex\hbox{$\sim$}\;$}}
\def\range{\hbox{$\,$\hbox{$-$}
	\kern-0.85em\raise.1ex\hbox{$:$} $\,$}}

\begin{document}

\title[Gravitational Wave Emission in CVs] 
{Gravitational Wave Emission by Cataclysmic Variables:\\ 
numerical models of semi-detached binaries.}

\author[Rezzolla, Ury\={u}, \& Yoshida ] 
{Luciano Rezzolla$^{1,2}$,~K\={o}ji Ury\={u}$^{1,3}$, 
\&  Shin'ichirou Yoshida$^1$. 
\\
$^1$SISSA, International School for Advanced Studies, 
Via Beirut 2-4, 34014 Trieste, Italy, \\
$^2$INFN, Department of Physics, University of Trieste,
Via Valerio 2, 34127 Trieste, Italy, \\
$^3$Department of Physics, University of Wisconsin-Milwaukee, 
P.O. Box 413, Milwaukee, WI 53201, U.S.A.}

\date{Received / Accepted}

\maketitle

\begin{abstract}
Gravitational wave emission is considered to be the
driving force for the evolution of short-period
cataclysmic binary stars, making them a potential test
for the validity of General Relativity. In spite of
continuous refinements of the physical description, a
10\% mismatch exists between the theoretical minimum
period ($P_{\rm turn} \simeq 70$ min) and the
short-period cut-off ($P_{\rm min} \simeq 80$ min)
observed in the period distribution for cataclysmic
variable binaries. A possible explanation for this
mismatch was associated with the use of the Roche
model. We here present a systematic comparison between
self-consistent, numerically constructed sequences of
hydrostatic models of binary stars and Roche models of
semi-detached binaries. On the basis of our approach, we
also derive a value for the minimum period of cataclysmic
variable binaries. The results obtained through the
comparison indicate that the Roche model is indeed very
good, with deviations from the numerical solution which
are of a few percent at most. Our results therefore
suggest that additional sources of angular momentum loss
or alternative explanations need to be considered in
order to justify the mismatch.
\end{abstract}


%%%%%%%%%%%%%%%%%%%%%%%%%%%%%%%%%%%%%%%%%%%%%%%

\section{Introduction}
\label{sect1}

	Semi-detached binary systems composed of a white
dwarf as the primary and a low-mass main sequence star or
brown dwarf as the secondary, are usually referred to as
Cataclysmic Variables, or CVs~\cite{w95}. Such systems
could show a cataclysmic behaviour as a consequence of
mass overflow from the secondary to the primary star. The
observed orbital period distribution of these systems has
two distinctive features: {\it (a)} a statistically
significant deficit of systems with orbital periods
between 2 and 3 hours, known as the ``period gap''; {\it
(b)} a sharp lower cutoff at an orbital period of $P_{\rm
min}\simeq 80$ min which is referred to as the ``minimum
period''~\cite{ritter-kolb}.

	In modeling short-period CVs (i.e. CVs with an
orbital period $< 2$~hr), it has long been proposed
that gravitational radiation reaction drives their
evolution enabling a continuous mass flow to occur from the
secondary to the primary
star~\cite{kraft-et-al,paczynski81,paczynski-sienkiewicz81,rappaport-et-al,dantona-mazzitelli,paczynski-sienkiewicz83,nelson-et-al,kolb-ritter,howell-et-al,kolb-baraffe}.
This feature makes CVs a potential test of General
Relativity~\cite{paczynski-sienkiewicz83}~\footnote{Note
that for CVs with an orbital period $> 3$~hr the emission
of gravitational waves cannot drive the binary evolution.}.

	The evolutionary scenario for a short-period CV
can be summarized as follows. Consider a binary system
with the secondary star having a mass 
%%$M_2 \gtrsim 0.1M_\odot$
$0.1M_\odot \lesssim M_2 \lesssim 0.3 M_\odot$\footnote{A
slightly different evolution scenario should be
considered in the case in which $M_2 \lesssim 0.1
M_\odot$~\cite{paczynski81}.}. As a result of the binary
evolution, the system becomes semi-detached, with the
secondary star filling its {\it ``critical lobe''} (which
is defined here as the three-dimensional equipotential
surface having a cusp at the inner Lagrangian
point)\footnote{We here introduce the term ``critical
lobe'' in order to distinguish between the critical
equipotential surface obtained from a self-consistent
three-dimensional solution of the hydrodynamical
equations and the corresponding surface obtained when
using the Roche model and which is usually referred to as
the ``Roche-lobe''.}. The loss of angular momentum then
induces a mass overflow from the inner Lagrange point
onto the primary degenerate dwarf. For the particular
class of CVs known as Dwarf Novae an accretion disc is
formed around the primary dwarf star and it is this which
is responsible for the observed ``cataclysmic''
behaviour.

	During the initial stages of the binary
evolution, the mass-transfer timescale $\tau_{\rm
{\dot{M}}}$, which is of the same order as the timescale
for angular momentum loss due to the gravitational wave
emission $\tau_{\rm GR}$, is much longer than the
Kelvin-Helmholtz timescale $\tau_{_{\rm KH}}$.
%i.e. $\tau_{\rm \dot{M}} \gg \tau_{_{\rm KH}}$. 
Under these conditions, the secondary can maintain its
thermal equilibrium and it slowly moves down the main
sequence as it loses mass. As a result of angular
momentum loss, the orbital period of the binary system
decreases and the mean density of the secondary
increases. A direct consequence of this is that
$\tau_{_{\rm KH}}$ for the secondary also increases. 

	 Because in the course of the evolution the ratio
$\tau_{\rm{\dot M}}/\tau_{\rm KH}$ decreases secularly,
the secondary gets more and more out of thermal
equilibrium. This is connected with a tendency of the
star to shrink less and less or even to expand upon mass
loss, depending on the value of $\tau_{\rm{\dot
M}}/\tau_{\rm KH}$.  The minimum (turning) period of the
binary orbit $P_{\rm turn}$ is reached for a specific
value of the ratio $\tau_{\rm{\dot M}}/\tau_{\rm KH}$
which depends somewhat on the model assumed for the
secondary [cf. eq. (2)]. After the minimum period is
reached, the nuclear burning becomes inefficient and
radiative losses from the surface of the star will reduce
its heat content. Eventually the star expands despite
losing mass and its evolutionary path moves away from the
main sequence towards the degenerate star branch.

	Within this evolutionary scenario, $P_{\rm turn}$
is usually determined by means of the so called Roche
model, which we will discuss in more detail in Section
\ref{rm}. The minimum period found by several authors in
this way~\cite{paczynski81,kolb-ritter,howell-et-al},
turns out to be about 10~\% shorter than the observed
period cut-off $P_{\rm min}$ (see Kolb \& Baraffe 1999
for the latest theoretical results). This difference
seems to survive all attempts at introducing refined
treatments of the stellar physics and of the stellar
evolution.

	There have been many proposals for solving this
discrepancy between $P_{\rm turn}$ and $P_{\rm min}$. One
of the suggestions is that the finite size effects of the
secondary star, that are neglected in the traditional
Roche model, may not be negligible and might be
responsible for the
mismatch~\cite{nelson-et-al,kolb-baraffe,lasota}.
Indeed, Nelson et al. (1985) have claimed that when tidal
distortion effects are included in their models on the
basis of the scheme suggested by Chan \& Chau (1979),
$P_{\rm turn}$ is increased by $\sim 10~\%$.  These
findings, however, were not confirmed by
subsequent investigations~\cite{kolb-baraffe}.

	In this paper, we focus our attention on the
validity of the Roche model for calculating $P_{\rm
turn}$ in CVs. For this doing, we compare the dynamical
properties of semi-detached binaries obtained by means of
the Roche model with those obtained by
numerically solving the equations of hydrostatic
equilibrium for a binary system. In our numerical
approach, we solve exact equations for the hydrostatic
equilibrium, so that the tidally deformed structure of 
the secondary star and the orbital motion of the binary
system are computed self-consistently. This is a feature
missing in the Roche model. In order to emphasize
the differences which emerge from the hydrodynamics of the
equilibrium binary system, our calculations do not
include details of the thermal structure or evolution of
the secondary.

	The paper is organized as follows: in Section
\ref{sect2} we describe the assumptions made in the
construction of the models for the binary system. In
Section \ref{sect3}, we discuss the set of equations that
is used in the numerical construction of the equilibrium
sequences of binary stars. Sections \ref{sect4} and
\ref{sect5} summarize the numerical results and present
our conclusions.

\section{Assumptions of the Models}
\label{sect2}

	A number of assumptions are necessary for the
construction of a model for a CV binary system. In
particular, following Paczy\'nski~ (1981), we consider a
semi-detached binary system with a white dwarf as the
primary star with mass $M_1$, and a hydrogen rich star on
the lower main sequence as the secondary star with mass
$M_2$.  $M_1$ and $M_2$ are assumed to be typically
$M_1\simeq 0.5-1.0M_\odot$ and $M_2<0.4M_\odot$. Since
the secondary is taken to be fully convective in this
mass range, its structure is well described by a
polytropic relation between the pressure $p$ and the mass
density $\rho$
\begin{equation}
\label{poly}
p=K\rho^{1+\frac{1}{N}} \ ,
\end{equation}
with polytropic index $N=3/2$ and polytropic coefficient
$K$, which is constant in space for an isentropic
model\footnote{Note that $K$ changes with $q$ along the
sequence of constant total mass equilibrium models.}. For
such a polytrope, the condition for the occurence of the
minimum period can then be expressed
as~\cite{dantona-mazzitelli-ritter,ritter96,kolb-ritter}
\begin{equation}
\label{taus}
\frac{\tau_{\rm \dot{M}}}{\tau_{_{\rm KH}}} \sim 0.3 \ .
\end{equation}
This condition, which contains some uncertainties related
to the stellar model assumed, will be used in Section
\ref{mtt} to determine the range of minimum periods from
the numerically calculated values of the mass transfer
timescale $\tau_{\rm \dot{M}}$.

	The radius of the primary is small compared to
that of the secondary and the primary is therefore
treated as a point-like mass. This simplifies the
calculation of both the gravitational potential of the
primary and its contribution to the quadrupole moment of
the system [cf. eq (\ref{I_ij})]. The binary orbit is
assumed to be circular and the spin period of the
secondary is taken to be synchronous with the orbital
period. These are rather good approximations since the
relaxation timescales to these states are much shorter
than the evolutionary timescale of the
system~\cite{zahn75,zahn77,lecar-et-al}.

	Finally, we assume that the secondary is always
just filling its critical lobe and that the continuous
mass-transfer from the secondary to the primary is
conservative, so that the total mass of the system $M =
M_1 + M_2$ is constant in time. This latter assumption is
justified when we consider the evolution of the binary
system to be driven solely by the emission of
gravitational radiation.

\section{Construction of the Models}
\label{sect3}

	In this section we briefly discuss the two
distinct sets of equations for the construction of models
for the CVs. In Section~\ref{rm}, we
summarize the set of equations describing the ``standard''
Roche model and in Section~\ref{nm}, we
discuss the set of the equations for a numerically
constructed hydrostatic model of a semi-detached binary.

\subsection{Roche model of a semi-detached binary}
\label{rm}

	We construct Roche models of CVs based on the
Roche model of semi-detached binaries as discussed by
Paczy\'nski in 1981. In this approximation, the secondary
star is assumed to fill up its Roche lobe and its
gravitational field is treated as that of a point-like
mass. Within this approach, it is convenient to measure
the size of the Roche lobe through a mean radius $r_{_L}$
defined so that the sphere of radius $r_{_L}$ has the
same volume $V_0$ as the Roche lobe
\begin{equation}
r_{_L} \equiv \left( \frac{3 V_0}{4\pi} \right)^{\frac{1}{3}}\ .
\end{equation}

	The ratio of $r_{_L}$ to the binary separation
$a$ depends only on the mass ratio $q\equiv M_2/M_1$ and
the empirical expressions proposed by Paczy\'nski (1971)
and Eggleton (1983) are expressed as
\begin{eqnarray}
\label{rocheR}
\frac{r_{_L}}{a} = f(q) = 
	\cases{
	\displaystyle
	\frac{2}{3^{\frac{4}{3}}}\left(\frac{q}{1+q}\right)^{\frac{1}{3}} 
		& (Paczy\'nski 1971)\cr
		&\cr
		&\cr
	\displaystyle
	\frac{0.49 q^{\frac{2}{3}}}{0.6 q^{\frac{2}{3}} + 
	\ln(1+q^{\frac{1}{3}})} 
		& (Eggleton 1983)\cr}\hskip -4truecm
\end{eqnarray}
The above expressions are valid for the range of mass
ratios of interest here: $q\simeq 0.1 - 1$.
Paczy\'nski's expression is accurate to within 1\% when compared
with the result obtained by numerical integration of the
Roche-lobe~\cite{kopal}. In the same range, Eggleton's
expression is even more accurate.

	Combining the above formula with Kepler's law
\begin{equation}
\label{kepler}
a^3 = \left(\frac{P}{2\pi}\right)^2 G (M_1 + M_2)\ ,
\end{equation}
we obtain an expression for the orbital period $P$ as 
\begin{equation}
\label{kepler2}
P =  \frac{2 \pi}{G^{\frac{1}{2}}} \left[
	\left(\frac{r_{_L}^3}{M_2}\right)
	\frac{q}{(1+q)(f(q))^3}\right]^{\frac{1}{2}}\ .
\end{equation}
Within the Roche model, the spin angular momentum
of the primary star is neglected and the total
angular momentum $J$ of the system is therefore given
by
\begin{equation}
\label{totJ}
J = J_{\rm orb} + J_{\rm spin},
\end{equation}
where the orbital angular momentum of the system $J_{\rm
orb}$ and the spin angular momentum of the secondary
$J_{\rm spin}$ are defined respectively as,
\begin{equation}
J_{\rm orb} = \left(G a \frac{M_1^2 M_2^2}{M_1+M_2}\right)^{\frac{1}{2}}
	= G^{\frac{1}{2}}
	\left[\frac{r_{_L}}{f(q)}\right]^{\frac{1}{2}}
	\left[\frac{M_2^3}{q(1+q)}\right]^{\frac{1}{2}}\ ,
\end{equation}
and
\begin{equation}
\label{Jspin}
J_{\rm spin} = (r_{g} r_L)^2 M_2 \Omega
 = (1+q)f^2(q) r^2_{g} J_{\rm orb}.
\end{equation}
Here $r_{g} r_L$ is the radius of gyration of the
secondary and for a polytrope with $N=3/2$, $r_{g} \simeq
0.45$~\cite{dantona-mazzitelli-ritter}. The theory of
stellar structure provides a simple mass--radius relation
which holds for lower main sequence stars in the mass
range of interest here (see, for example, Kippenhahn \&
Weigert 1990). This relation can be written as
\begin{equation}
\label{r_vs_m}
\frac{R_2}{R_\odot} = \chi \left(\frac{M_2}{M_\odot}\right)^{\xi}\ ,
\label{masradi}
\end{equation}
where $M_\odot$ and $R_\odot$ are the solar mass and the
solar radius, respectively, and $R_2$ is the radius of
the lower main sequence star. For a star with chemical
abundances similar to those of the Sun, we can set
$\chi=0.83$ and $\xi=0.84$ (Kippenhahn \& Weigert
1990). Moreover, as long as the star remains on the lower
main sequence (i.e. as long as $\tau_{_{\rm KH}} < \tau_{\rm
{\dot{M}}}$), we can assume that the mass--radius
relation (\ref{r_vs_m}) holds 
even in the presence of mass loss, i.e. when
\begin{equation}
\label{roapp}
r_{_L}=R_2 \ .
\end{equation}
Using equations (\ref{masradi}) and (\ref{roapp}), we
can now rewrite the expression for the orbital period
(\ref{kepler2}) as
\begin{equation}
\label{p_vs_rho}
P = 2.78 \times
	\left[\chi^{3} \left(\frac{M_2}{M_\odot}\right)^{3\xi-1}
	\frac{q}{(1+q)(f(q))^3}\right]^{\frac{1}{2}}  {\rm hr} \ ,
\end{equation}
and derive the two timescales which are relevant
for the evolution of a CV binary: namely the timescale for
gravitational wave emission $\tau_{\rm GR}$ and that
for mass-transfer $\tau_{\rm {\dot{M}}}$. As
mentioned above, we assume that the binary evolution is
driven only by the emission of gravitational waves
(GW) and the timescale for this to occur can be estimated
by means of the standard quadrupole formula for two point-like
masses orbiting around each other~\cite{landau-lifshitz}
\begin{eqnarray}
\label{quadrupole}
&& \hskip -0.65 truecm
	\tau_{_{\rm GW}} \equiv \left(\left.\frac{d\ln J}{dt}
	\right|_{\rm GW}\right)^{-1} 
\nonumber\\	
	&& \hskip 0.05 truecm
	= -\frac{5}{32}
	\left(\frac{P}{2\pi}\right)^{\frac{8}{3}}\frac{c^5}{G^{\frac{5}{3}}}
        \frac{(M_1+M_2)^{\frac{1}{3}}}{M_1M_2} 
\nonumber\\
	&& \hskip 0.05 truecm
        = -7.87\times 10^{7} 
	\left(\frac{P}{1\ {\rm hr}}\right)^{\frac{8}{3}}
	\left(\frac{M_2}{M_\odot}\right)^{-\frac{5}{3}} 
	\left[{q^2(1+q)}\right]^{\frac{1}{3}} \ {\rm yr}
	\ ,
\nonumber\\
\end{eqnarray}
On the other hand, we can calculate the mass-transfer
timescale as 
\begin{equation}
\label{taumdot}
\tau_{\rm {\dot{M}}} \equiv 
	-\left(\frac{d\ln M_2}{dt} \right)^{-1}
	=-\left[\frac{d\ln M_2}{d\ln J} 
	\left.\frac{d\ln J}{dt}\right|_{\rm
	GW}\right]^{-1} \ , 
\end{equation}
where the logarithmic differential of the total angular
momentum (\ref{totJ}) can be expressed as
\begin{eqnarray}
d\ln J &=& \frac{1}{2}d\ln r_{_L} + \frac{3}{2} d\ln M_2
\nonumber \\  
	&&\hskip  1.5 truecm 
	- \frac{1}{2}\left[ \frac{d\ln f(q)}{dq} + 
	  \frac{2q+1}{q(1+q)} \right] dq 
 \nonumber \\ 
       &=& \frac{1}{2}d\ln M_2
	\left[2 - 2q - q(1+q)\frac{d}{dq}\ln f(q) + \xi
	\right] , 
\label{dlogJ}
\end{eqnarray}
where we have used equations (\ref{masradi}) and
(\ref{roapp}) as well as the conservation of the total
mass of the system
\begin{equation}
dq = q(q+1) d\ln M_2\ .
\end{equation}
Combining equations (\ref{quadrupole}) -- (\ref{dlogJ}), we can
write explicitly the mass-transfer timescale for the
secondary as
\begin{eqnarray}
\tau_{\rm {\dot{M}}} 	&=&  - \frac{1}{2}
		\left(\left.\frac{d\ln J}{dt}\right|_{\rm GW}\right)^{-1}
	\nonumber\\
	&& \hskip 1.0 truecm
	\times \left[2 - 2q - q(1+q)\frac{d}{dq}\ln f(q) + \xi
		\right]\ .
\label{tauMroche}
\end{eqnarray}
Note that the CV system described by the Roche model is
determined in terms of two variables: the mass ratio $q$
and the mass of the secondary, $M_2$.

\subsection{Numerical hydrostatic model of a semi-detached binary}
\label{nm}

	A numerical method for computing semi-detached
compressible binaries in hydrostatic equilibrium has 
recently been developed by Ury\={u} \& Eriguchi (1999) [see
also Ury\={u} \& Eriguchi~(1998) for the case of
incompressible binaries.]. Within the assumptions
discussed in Section \ref{sect2}, and in a corotating
reference frame, the matter of the secondary star satisfies the
following equation of hydrostatic equilibrium (the
Bernoulli equation)
\begin{equation}
\label{bernou}
-\frac{1}{2}\varpi^2\Omega^2 + K(N+1)\Theta + \phi = C \ ,
\end{equation}
where, $\phi$ and $\Omega$ are the gravitational
potential, and the angular velocity of the orbital
motion, respectively.  $\varpi$ is the distance of the
fluid element from the rotation axis of the binary and
$C$ is a constant. The function $\Theta$ is defined by
$\rho=\Theta^{N}$. The gravitational potential
$\phi$, is computed using the integral expression
\begin{equation}
\label{phi}
\phi(\mbox{\bf r}) 
	= -G\int_{V_2}
	\frac{\rho(\mbox{\bf r}')}{|\mbox{\bf r}-\mbox{\bf r}'|}dV
	-\frac{G M_1}{|\mbox{\bf r}-\mbox{\bf r}_1|}\ ,
\end{equation}
where the volume integration is carried out in the
interior of the secondary ($V_2$), and $\mbox{\bf r}_1$
is the position vector of the primary. Equations
(\ref{bernou}) and the integral in the equation
(\ref{phi}) are discretized in a spherical coordinate
system $(r,\theta,\varphi)$ whose origin is at the centre
of the secondary. Because of the symmetry of the
configuration, we only compute the solution in the region
$r\in [0,R_0],\; \theta\in [0,\pi/2],\; \varphi\in
[0,\pi]$, where $R_0$ is the maximum radius of a tidally
deformed stellar surface. A solution is obtained on a
grid with the number of gridpoints for each coordinate
being $(N_r,N_\theta,N_\varphi) = (65,33,65)$. The
Green's function of the integral in the equation
(\ref{phi}) is expanded in Legendre polynomials
$P_l(\cos\theta)$ up to $\ell=14$. The number of grid
points as well as the number of Legendre polynomials used
for the Green's function are chosen so as to provide
satisfactory numerical accuracy and results presented in
this paper have a numerical error which is typically less
than $0.1\%$. [Further details of the numerical method
can be found in Ury\={u} \& Eriguchi (1999) and
references therein.]

	In calculating models of ``critical
lobe-filling binaries'' (hereafter critical lobe
binaries), we first fix a value for the mass ratio $q$
and then compute a set of solutions corresponding to different
values of the orbital separation $a$ (we start from a
large initial value of $a$ and then decrease it in
steps of $\delta a = R_0/50$). The sequence of solutions
obtained in this way is terminated at an orbital
separation for which the surface of the secondary star has
a cusp at the inner Lagrange point. The corresponding
configuration is identified as a critical lobe model for
the given value of $q$. This procedure has been repeated
for 31 models, changing $q$ logarithmically in the range
$q \in [0.033,\; 1]$.

	For each of the 31 models, we compute the
relevant timescales and, as in equation
(\ref{quadrupole}), $\tau_{_{\rm GW}}$ is given by the
rate of angular momentum loss from the system in the
form of gravitational radiation~\cite{mtw},
\begin{equation}
\label{dJdtGR}
\left.\frac{d\ln J}{dt}\right|_{\rm GW}
	= -\frac{32G}{5c^5}\frac{\Omega^5}{J}(I_{11}-I_{22})^2
	\ .
\end{equation}
The quadrupole moment tensor $I_{ij}$ of the system
is computed numerically as 
\begin{equation}
\label{I_ij}
I_{ij} \equiv \int_V (\rho_{_S} + \rho_{_P}) 
	\left(x_i x_j	- \delta_{ij}
	\frac{|\mbox{\bf r}|^2}{3}\right) dV\ ,
\end{equation}
where the volume $V$ comprises both stars and
$\rho_{_S},\;\rho_{_P}$ are the mass densities of the
secondary and primary stars, respectively, with the
latter being expressed in terms of a three-dimensional
Dirac delta function $\rho_{_P} \equiv M_1 \delta^3({\bf r} -
\varpi_1 {\bf e}_{x_1})$. The vector $\mbox{\bf
r}=(x_1,x_2,x_3)$ is the position vector of a mass
element and $\varpi_1$ represents the distance of the
primary star (a point particle) from the rotation centre
of the binary. The vector ${\bf e}_{x_1}$ is the unit vector
along the $x_1$ axis. The Latin indices in (\ref{I_ij}) run from
1 to 3 and correspond to the components of the Cartesian
coordinate system whose origin is at the rotation centre
of the binary system (the $x_1-$coordinate is oriented along the
separation vector of the binary and the $x_2-$coordinate
is in the orbital plane). The total angular momentum $J$
for the model is computed as follows :
\begin{equation}
J = \int_{V_2} {\rho \varpi^2 \Omega dV} + M_1 \varpi_1^2 \Omega \ .
\label{Jtot}
\end{equation}
As in equation (\ref{taumdot}), we calculate the
mass-transfer timescale as
\begin{eqnarray}
{\tau_{\rm {\dot{M}}}} \equiv - \left( \frac{d\ln M_2}{dt} \right)^{-1}
	= \left(\frac{d\ln M_2}{d\ln J}
	\left.\frac{d\ln J}{dt}\right|_{\rm GW}\right)^{-1}\ ,
\label{tauMnum}
\end{eqnarray}
where now the quantity $d\ln M_2/d\ln J$ is computed
along the semi-detached equilibrium sequence from the
expression
\begin{equation}
\label{dmdj}
\frac{d\ln M_2}{d\ln J} = \frac{1}{q(q+1)}\frac{dq}{d\ln J}\ .
\end{equation}

\section{Comparison of the Models}
\label{sect4}

	In this Section, we compare the two classes of
semi-detached binary models, namely the Roche model and
the numerical critical lobe model in hydrostatic
equilibrium. We first compare purely dynamical quantities
for the two models. We then discuss the differences
arising from the application of these two models to a
cataclysmic binary system, focusing in particular on the
differences in the mass-transfer timescale.

\begin{figure}
\centering\leavevmode
\psfig{file=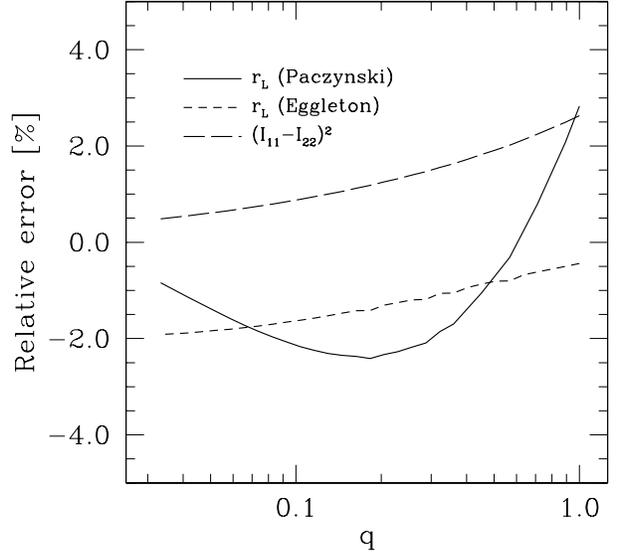,width=8.4cm,angle=0,clip=}
\caption{The relative differences of several quantities
when computed using the Roche model and the numerical
binary models are plotted against $q$. The solid line is
the result for $r_{_L}$ as computed with Paczy\'nski's
formula. The short-dashed line is the same but for
Eggleton's formula. The long-dashed line, 
is the relative error for $(I_{11}-I_{22})^2$ [see
eq. (\ref{I_ij})].  }
\label{fig1}
\end{figure}
\begin{figure}
\centering\leavevmode
\psfig{file=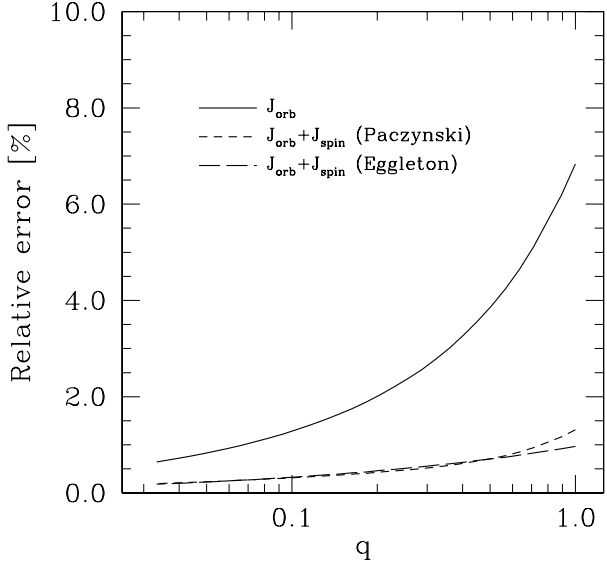,width=8.4cm,angle=0,clip=}
\caption{The relative deviations of the total angular
momentum when computed using the Roche model and the
numerical binary models are plotted against $q$. The
solid line refers to binaries in which the spin angular
momentum of the secondary is neglected in the Roche
model. Binaries in which the spin contribution of
secondary in the Roche model is instead included are
indicated with a short-dashed line (for the case of
Paczy\'nski's formula) and with a long-dashed line (for
the case of Eggleton's formula). }
\label{fig2}
\end{figure}

\subsection{Comparison of dynamical properties}

	Two binaries are considered to represent the same
physical system (and are therefore considered
``comparable'') if they have the same mass ratio $q$ and
the same orbital period $P$. In practice, we first
construct a numerical model of a critical lobe binary and
then calculate the physical quantities of Roche model
binary with the same mass ratio and orbital period. Note
that the relation (\ref{masradi}) between the mean radius
and the mass of the secondary is not used in this
comparison, but we match two different binaries on the
basis of the dynamical timescale of the system (i.e. the
orbital period of the binary system).

\begin{figure}
\centering\leavevmode
\psfig{file=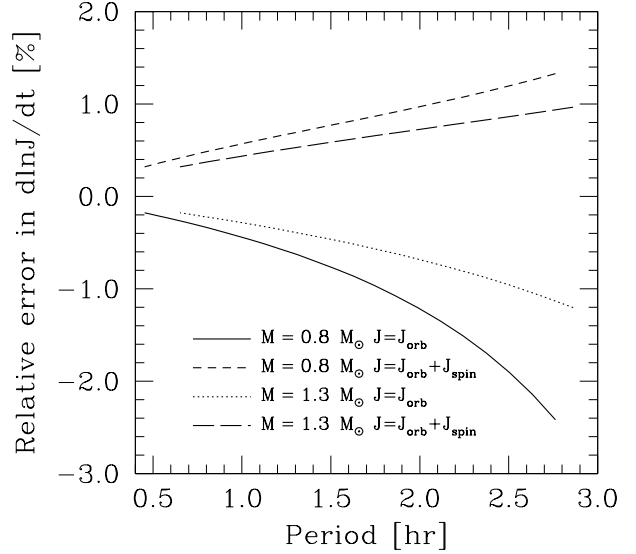,width=8.4cm,angle=0,clip=}
\caption{Relative error of the inverse of the timescale
for angular momentum loss by gravitational radiation,
$d\ln J/dt$. The solid line refers to a binary system
with total mass $M=0.8M_{\odot}$ and does not include the
spin angular momentum contribution of the secondary,
which is instead included in the curve indicated with a
short-dashed line. Similarly, the long-dashed and dotted
lines are for a system having a total mass
$M=1.3M_{\odot}$, with and without the inclusion of the
spin angular momentum of the secondary.}
\label{fig3}
\end{figure}

	In the comparison we focus on: {\it (a)} the mean
radius of the secondary $r_{_L}$; {\it (b)} the total
angular momentum of the system $J$; {\it (c)} the
components of the quadrupole moment tensor
$(I_{11}-I_{22})^2$ and {\it (d)} the inverse of the
timescale for gravitational wave emission $d \ln
J/dt$. For all of these quantities we define the measure
of the relative deviation as
\begin{eqnarray}
\mbox{Relative error in } Q \equiv
	\frac{Q(\mbox{critical lobe})-Q(\mbox{Roche model})}
	{Q(\mbox{Roche model)}} \ , &&
	\nonumber \\
\end{eqnarray}
and have plotted them in Figures \ref{fig1}-\ref{fig2}
against the mass ratio of the system, and in Figure
\ref{fig3} against the orbital period.

	In particular, the solid line in Figure
\ref{fig1} shows the mean radius of the critical lobe,
$r_{_L}$, from Paczy\'nski's formula, while the
short-dashed line comes from using Eggleton's
formula. The long-dashed line, on the other hand, shows
the relative error for $(I_{11}-I_{22})^2$\footnote{We
recall that in the Roche model $I_{11}-I_{22}= a^2
M_1M_2/(M_1+M_2)$.}. Similarly, Figure \ref{fig2} shows
the relative deviations of the total angular momentum:
the solid line refers to binaries in which the spin
angular momentum of secondary is neglected in the Roche
model, while binaries for which the spin contribution of
secondary in Roche model is included are indicated with
short and long-dashed lines. Both Figures \ref{fig1} and
\ref{fig2} show that for all of these quantities the
errors made through the use of the Roche model are
smaller than 3\%, if the spin angular momentum of the
secondary is taken into account.

	In a similar way, in Figure \ref{fig3}, where we
calculate the relative error in $d\ln J/dt$ for different
values of the orbital period, the different line types
refer to two systems having a total mass of
$M=0.8M_{\odot}$ and $M=1.3M_{\odot}$, respectively. In
this comparison we have used Eggleton's formula but we
have verified that the error is almost the same when
Paczy\'nski's formula is used.
	Also in this case, the deviations are of at most
a few percent, as found by Kolb \& Baraffe (1999) when
implementing the prescription of Chan \& Chau (1979) for
the corrections due to tidal distortion.

	Note that in evaluating the timescale for the
angular momentum loss, the accuracy of the Roche model
increases as the mass of the secondary is decreased and
thus for shorter orbital periods. This can be deduced
from the fact that, for a fixed value of the total mass
$M$, the relative error in $d\ln J/dt$ becomes smaller
for smaller orbital periods and therefore for smaller
masses of the secondary. The improved accuracy is related
to the fact that the Roche model is increasingly accurate
as the finite-size effects become less and less
important, or equivalently as the ratio $r_{_L}/a$
becomes smaller and smaller. In the case of Paczy\'nski's
formula, this ratio can be expressed as
\begin{equation}
\frac{r_{_L}}{a} = \frac{2}{3^{\frac{4}{3}}}
	\left(\frac{M_2}{M}\right)^{\frac{1}{3}} \ .
\end{equation}
showing that for a fixed total mass $M$, finite-size
effects will be smaller with a less massive secondary star.

\subsection{Mass-transfer timescale}
\label{mtt}

\begin{figure}
\centering\leavevmode
\psfig{file=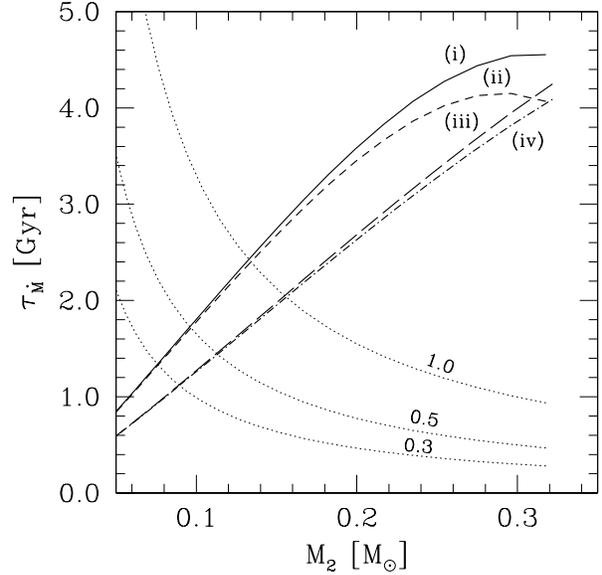,width=8.4cm,angle=0,clip=}
\caption{$\tau_{\rm {\dot{M}}}$ is plotted as a function
of the mass of the secondary.  Curves (i) and (ii) are for a
binary system with a total mass of $0.8M_{\odot}$. Curve
(i) corresponds to the result of the numerical
computation, whereas (ii) is obtained with the Roche model
based on the Eggleton formula.  Curves (iii) and (iv)
are similar to (i) and (ii), but for a system with total
mass $1.3M_{\odot}$. Also plotted are the curves of
${\tilde \tau}_{_{\rm KH}}$ for the secondary. The labels
of these curves refer to the values of the parameter
$\alpha$ defined in equation (\ref{tau_th}). }
\label{fig4}
\end{figure}

	Next, we compare the mass-transfer timescale
$\tau_{\rm {\dot{M}}}$ computed from equations
(\ref{quadrupole}) and (\ref{tauMroche}) for the Roche
model binaries, and equations
(\ref{dJdtGR})--(\ref{dmdj}) for the numerical
critical-lobe binaries. For the Roche model, we use
Eggleton's formula in equation (\ref{rocheR}).  Note that
the mass-radius relation (\ref{masradi}) is used to
compute $\tau_{\rm {\dot{M}}}$.  In Figure \ref{fig4},
curves (i) and (ii) refer to a binary system with a total
mass of $0.8M_{\odot}$, with the first one being the
result of the fully numerical calculation and the second
one being obtained using the Roche model\footnote{For
simplicity, we do not include here the spin angular
momentum contribution of the secondary in the Roche
model. As shown in Figure \ref{fig3}, this produces an
additional error of a few percent at most.}.  Curves
(iii) and (iv) are similar to (i) and (ii), but refer to
a system with total mass $1.3 M_{\odot}$.
	
	An important aspect emerging from Figure
\ref{fig4} is that, for all of the relevant masses of the
secondary, the mass-transfer timescale computed
numerically is slightly {\it larger} than that obtained
from the Roche model.  In other words, the Roche model
slightly {\it underestimates} the evolutionary timescale,
thus resulting in a {\it larger} minimum orbital
period. As a consequence, our results indicate that a
treatment which refines the hydrodynamic aspects of the
critical lobe model for CVs produces even larger
discrepancies with the observations.

	In Figure \ref{fig4}, we also show the curves for
a parametrized timescale ${\tilde \tau}_{_{\rm KH}}$ 
\begin{equation}
\label{tau_th}
{\tilde \tau}_{_{\rm KH}} \equiv \alpha \tau_{_{\rm KH}} =
\alpha \frac{GM_2^2}{R_2 L_2}\ ,
\end{equation}
where the adjustable coefficient $\alpha \in [0.25, 1.0]$
has been introduced to account for the uncertainty in the
determination of the minimum period
[cf. eq. (\ref{taus})]

	For a low-mass main sequence star, the luminosity
of the secondary $L_2$ appearing in (\ref{tau_th}) can be
expressed as~\cite{kippenhahn-weigert},
\begin{equation}
\frac{L_2}{L_\odot} = 0.1413 \left(
	\frac{M_2}{M_\odot} \right)^{2.25}\ .
\end{equation}
The orbital period at which $\tau_{\rm {\dot{M}}}=
{\tilde \tau}_{_{\rm KH}}$ corresponds to the minimum
period of the binary system $P_{\rm turn}$.  In Tables I
and II, we show the values of the binary parameters and
the explicit values of $P_{\rm turn}$ at $\tau_{\rm
{\dot{M}}}={\tilde \tau}_{_{\rm KH}}$ for several values
of $\alpha$ and $M$.

%--- Fig4
%---         

%--- Table 1---
\begin{table}
\begin{center}
\begin{tabular}{ccccccc}
\hline
\\[-2.5mm]
& model & $M_2/M_\odot$ & $q$ & $\tau_{\rm {\dot{M}}} $ [Gyr] & $P_{\rm turn}$ [min] & 
\\[0.5mm]
\hline
\multicolumn{7}{c}{$\alpha = 0.25$} \\
\hline
& Roche & 0.0824 & 15.05 & 1.019 & 60.20 & \\
& Numerical  & 0.0821 & 15.11 & 1.025 & 60.02 & \\
\\[-3mm]
\hline
\multicolumn{7}{c}{$\alpha = 0.3$} \\
\hline
& Roche & 0.0896 & 13.77 & 1.1174 & 64.28 & \\
& Numerical  & 0.0893 & 13.83 & 1.119 & 64.08 & \\
\\[-3mm]
\hline
\multicolumn{7}{c}{$\alpha = 0.35$} \\
\hline
& Roche & 0.0962 & 12.77 & 1.208 & 67.96 & \\
& Numerical  & 0.0958 & 12.83 & 1.204 & 67.74 & \\
\\[-3mm]
\hline
\multicolumn{7}{c}{$\alpha = 0.5$} \\
\hline
& Roche & 0.1133 & 10.68 & 1.442 & 77.22 & \\
& Numerical  & 0.1127 & 10.74 & 1.449 & 76.92 & \\
\\[-3mm]
\hline
\multicolumn{7}{c}{$\alpha = 1.0$} \\
\hline
& Roche & 0.1561 & 7.483 & 2.034 & 99.12 & \\
& Numerical  & 0.1550 & 7.540 & 2.049 & 98.58 & \\
\\[-3mm]
\\[-3mm]
\hline
\end{tabular}
\end{center}
\caption{ Comparison of binary parameters at the point along the
evolutionary sequence where the relation, $\tau_{\rm
{\dot{M}}} = {\tilde \tau}_{_{\rm KH}}$, hold. 
The total mass of the system is $1.3
M_{\odot}$.
\label{table1}
}
\end{table}
%-----------------
%\hline
%\multicolumn{7}{c}{$\alpha = 1.5$} \\
%\hline
%& Roche & 0.1887 & 6.014 & 2.480 & 114.7 & \\
%& Numerical  & 0.1871 & 6.071 & 2.502 & 114.0 & \\
%\\[-3mm]

%--- Table 2-------
\begin{table}
\begin{center}
\begin{tabular}{ccccccc}
\hline
\\[-2.5mm]
& model & $M_2/M_\odot$ & $q$ & $\tau_{\rm {\dot{M}}} $ [Gyr] 
& $P_{\rm turn}$ [min] & 
\\[0.5mm]
\hline
\multicolumn{7}{c}{$\alpha = 0.25$} \\
\hline
& Roche & 0.0702 & 10.67 & 1.215 & 53.66 & \\
& Numerical  & 0.0698 & 10.72 & 1.201 & 53.45 & \\
\\[-3mm]
\hline
\multicolumn{7}{c}{$\alpha = 0.3$} \\
\hline
& Roche & 0.0763 & 9.725 & 1.330 & 57.28 & \\
& Numerical  & 0.0759 & 9.785 & 1.311 & 57.05 & \\
\\[-3mm]
\hline
\multicolumn{7}{c}{$\alpha = 0.35$} \\
\hline
& Roche & 0.0820 & 8.997 & 1.437 & 60.58 & \\
& Numerical  & 0.0815 & 9.055 & 1.412 & 60.31 & \\
\\[-3mm]
\hline
\multicolumn{7}{c}{$\alpha = 0.5$} \\
\hline
& Roche & 0.0967 & 7.471 & 1.714 & 68.88 & \\
& Numerical  & 0.0960 & 7.528 & 1.727 & 68.52 & \\
\\[-3mm]
\hline
\multicolumn{7}{c}{$\alpha = 1.0$} \\
\hline
& Roche & 0.1341 & 5.109 & 2.401 & 88.56 & \\
& Numerical  & 0.1328 & 5.172 & 2.427 & 87.90 & \\
\\[-3mm]
\hline
\end{tabular}
\end{center}
\caption{
The same as Table 1, except that the total mass of the system is
$0.8M_{\odot}$
\label{2}
}
\end{table}
%-----------------
%\hline
%\multicolumn{7}{c}{$\alpha = 1.5$} 		\\
%\hline
%& Roche & 0.1634 & 4.012 & 2.902 & 102.9 & 	\\
%& Numerical  & 0.1613 & 4.080 & 2.945 & 101.9 & \\
%\\[-3mm]

\section{Conclusions}
\label{sect5}

	The Roche model has been widely used in the
literature to describe the dynamical evolution of
CVs. The minimum periods computed on the basis of this
model have shown a disagreement with observations and
several attempts made to include more realistic
descriptions of the system have not yet produced a
satisfactory explanation for it. In order to assess the
accuracy of the Roche model we have carried out numerical
calculations of equilibria of corotating semi-detached
binaries.  The models obtained numerically are ``exact''
in the sense that they are hydrodynamically
self-consistent and fully account for the finite size of
the secondary star which fills its critical lobe.

	However, when physical systems obtained with the
two approaches are compared, these show only minimal
differences which are always smaller than a few percent.
Therefore, our results indicate that the explanation for
the mismatch between the theoretical minimum period and
the observed one must be due to something else.  A better
agreement with observations might be found when
mechanisms that increase the efficiency of angular
momentum loss from the system, are fully taken into
account. Among these mechanisms one could consider, for
instance, stellar winds or magnetic torques, both of
which have been shown to be very efficient in removing
angular momentum. An interesting alternative mechanism
which could increase the angular momentum loss via
gravitational waves is given by the possible excitation
of stellar oscillations in binary systems. These periodic
perturbations, which could be excited in binary systems
such as CVs, would generate gravitational wave emission
which would add to the one coming from the orbital motion
and, when resonant, could even be the largest source of
gravitational radiation~\cite{bf00}.

\section*{Acknowledgments}

We are grateful to Jean-Pierre Lasota for suggesting this
research and for his useful comments, and to John Miller
for carefully reading the manuscript. We also thank the
referee, Hans Ritter, for his detailed
suggestions. Financial support for this research has been
provided by the MURST and by the EU Programme "Improving
the Human Research Potential and the Socio-Economic
Knowledge Base" (Research Training Network Contract
HPRN-CT-2000-00137).

%%%%%%%%%%%%%%%%%%%%%%%%%%%%%%%%%%%%%%%%%%%%%%%%

\end{document}